\documentclass[multphys,vecphys]{svmult}

\usepackage{makeidx}         
\usepackage{graphicx}        
\usepackage{multicol}        
\usepackage[bottom]{footmisc}

\makeindex             

\begin{document}

\title*{An Origin of CMR: Competing Phases and Disorder-Induced 
Insulator-to-Metal Transition in Manganites}
\titlerunning{An Origin of CMR}
\author{Yukitoshi Motome\inst{1}\and 
Nobuo Furukawa\inst{2}\and Naoto Nagaosa\inst{3}\inst{4}\inst{5}}
\institute{RIKEN (The Institute of Physical and Chemical Research), 
2-1 Hirosawa, Saitama 351-0198, Japan
\texttt{motome@riken.jp}
\and Department of Physics, Aoyama Gakuin University, 
5-10-1 Fuchinobe, Sagamihara, Kanagawa 229-8558, Japan
\texttt{furukawa@phys.aoyama.ac.jp}
\and CREST, Department of Applied Physics, University of Tokyo,
7-3-1 Hongo, Bunkyo-ku, Tokyo 113-8656, Japan
\texttt{nagaosa@appi.t.u-tokyo.ac.jp}
\and Correlated Electron Research Center, AIST,
Tsukuba Central 4, 1-1-1 Higashi, Tsukuba, Ibaraki 305-8562, Japan
\and Tokura Spin SuperStructure Project, ERATO,
Japan Science and Technology Corporation, c/o AIST, 
Tsukuba Central 4, 1-1-1 Higashi, Tsukuba, Ibaraki 305-8562, Japan
}
%
%
\maketitle

We theoretically explore the mechanism of the colossal magnetoresistance 
in manganese oxides by explicitly taking into account 
the phase competition between the double-exchange ferromagnetism and 
the charge-ordered insulator.
We find that quenched disorder 
causes a drastic change of the multicritical phase diagram 
by destroying the charge-ordered state selectively. 
As a result, there appears a nontrivial phenomenon of the disorder-induced 
insulator-to-metal transition in the multicritical regime. 
On the contrary, the disorder induces a highly-insulating state 
above the transition temperature
where charge-ordering fluctuations are much enhanced. 
The contrasting effects provide an understanding of 
the mechanism of the colossal magnetoresistance. 
The obtained scenario is discussed in comparison with 
other theoretical proposals such as 
the polaron theory, the Anderson localization, 
the multicritical-fluctuation scenario, and 
the percolation scenario.

\section{Introduction}
\label{sec:1}

Colossal magnetoresistance (CMR) in perovskite manganese oxides $A$MnO$_3$ 
has attracted much attention in the physics of 
strongly-correlated electron systems
\cite{Ramirez1997,Tokura2000,Dagotto2001}. 
In these materials, carrier doping by chemical substitutions of $A$-site ions, 
e.g., Ca substitutions into La-site in LaMnO$_3$, 
leads to a ferromagnetic metal (FM) at low temperatures. 
In this regime, the resistivity shows a rapid decrease 
by applying an external magnetic field near the critical point, 
which is called the CMR effect. 
The doped FM state and the negative magnetoresistance are basically 
understood by the Zener's double-exchange (DE) interaction
\cite{Zener1951,Furukawa1999}. 
In the Zener's scenario, the system consists of 
conduction electrons and localized spins, and 
there is a strong ferromagnetic Hund's coupling between them. 
Through this strong correlation, 
the external magnetic field which aligns the localized spins ferromagnetically 
increases the kinetic energy of electrons to induce a metallic state. 

Recent development in experiments, partly promoted by 
potential applications to electronic engineering 
such as spintronics devices, 
has achieved to enhance the CMR effect; 
the resistivity sharply decreases on the order of $10^4-10^6$ 
by the magnetic field of only a few Tesla. 
One of the important features of the enhanced CMR is 
a characteristic temperature dependence of the resistivity 
at zero magnetic field. 
The resistivity shows good metallic behavior below 
the ferromagnetic transition temperature $T_{\rm C}$ 
while it shows highly-insulating behavior above $T_{\rm C}$. 
That is, the resistivity shows a steep increase toward $T_{\rm C}$ from above 
and suddenly drops near $T_{\rm C}$. 
The highly-insulating state just above $T_{\rm C}$ 
is very sensitive to the magnetic field, 
which leads to the enhanced CMR effect. 
Therefore, the origin of the highly-insulating state above $T_{\rm C}$ 
is a key to understand the mechanism of the enhanced CMR effect. 

Either the highly-insulating nature or the huge response 
to the external magnetic field cannot be explained by the simple DE theory. 
There have been many theoretical proposals which attempt to explain them. 
Several scenarios are based on the single-particle picture 
such as spin polaron 
\cite{Varma1996}
or Jahn-Teller (JT) polaron theory 
\cite{Millis1995,Roder1996}
and Anderson localization scenario by quenched disorder
\cite{Letfulov2001,Narimanov2002}. 
Recently, some attempts have also been made 
to understand the enhanced CMR as a cooperative phenomenon 
due to the many-body correlation, 
such as the multicritical-fluctuation scenario 
\cite{Murakami2003}
and the percolation scenario 
\cite{Moreo1999,Burgy2001}. 
It is highly desired to clarify which is a suitable picture. 

Several experiments indicate that 
the importance of a keen competition between different phases 
and quenched disorder for the CMR phenomena. 
One of the systematic investigations has been explored 
in a new class of materials 
$A_{1/2}$Ba$_{1/2}$MnO$_3$ 
\cite{Millange1998,Nakajima2002,Akahoshi2003}. 
It is found that under a special condition of the synthesis, 
$A$ ions and Ba ions constitute a periodic layered structure. 
In these $A$-site ordered materials, 
every Mn site has an equivalent environment of the surrounding ions, 
and therefore, there is no disorder from the alloying effect. 
In a usual synthesis process or by annealing the $A$-site ordered materials, 
one obtains the materials in which $A$ and Ba ions 
distribute randomly, and there is structural and electrostatic 
disorder at Mn sites. 
Since the systematic change of the average radius of $A$-site ions 
is known to modify the electron bandwidth, 
it is now possible to control 
both electron correlation and quenched disorder 
in a systematic way in these new materials.

\begin{figure}
\centering
\includegraphics[width=8cm]{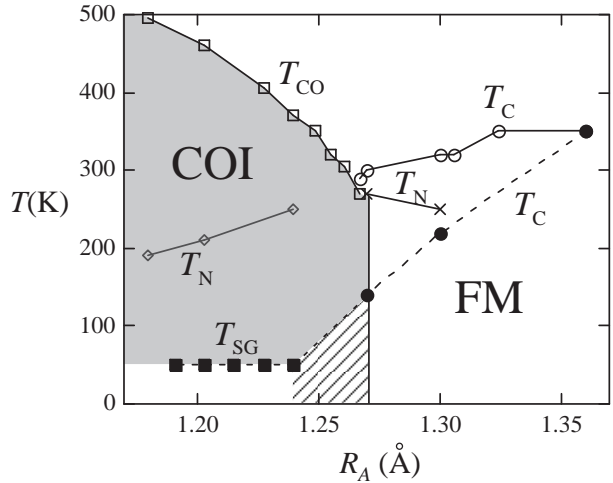}
\caption
{
Experimental phase diagram for $A_{1/2}$Ba$_{1/2}$MnO$_3$ 
obtained in Ref.~\cite{Akahoshi2003}. 
The horizontal axis is the averaged ionic radius of the $A$-site ion. 
$T_{\rm C}$, $T_{\rm CO}$, $T_{\rm N}$, and $T_{\rm SG}$ 
denote the transition temperatures for 
FM, COI, the antiferromagnetic state, and the glassy state, respectively. 
The hatched area is the disorder-induced COI-to-FM transition regime. 
The shaded area shows the region where COI is destroyed 
by introducing the $A$-site disorder. 
See the text and Ref.~\cite{Akahoshi2003} for details. 
}
\label{fig:phase diagram exp}
\end{figure}

Figure~\ref{fig:phase diagram exp} shows the phase diagram 
of this class of materials obtained by Akahoshi {\it et al.}
\cite{Akahoshi2003}. 
Open symbols connected by the solid line in the figure are 
the results in the $A$-site ordered materials. 
There, the system shows a typical multicritical phase diagram, 
that is, the FM transition temperature $T_{\rm C}$ and 
the charge-ordered insulator (COI) transition temperature $T_{\rm CO}$ 
meet with each other at almost the same temperature 
(the multicritical point). 
In the disordered materials, 
the multicritical phase diagram shows a drastic change 
in an asymmetric manner as shown by 
the closed symbols connected by the dashed line in the figure. 
That is, the FM state is robust in spite of 
a suppression of $T_{\rm C}$, 
whereas the COI state completely disappears and is replaced 
by a glassy state at very low temperatures below $T_{\rm SG}$. 
The multicritical point is shifted to 
the left side of the phase diagram, 
and at the same time, is suppressed down to a lower temperature. 
Consequently, there appears a nontrivial regime
in which the disorder induces the transition from COI to FM 
(the hatched area in Fig.~\ref{fig:phase diagram exp}). 

An important observation is that 
in this regime of the disorder-induced COI-to-FM transition, 
the resistivity shows the characteristic temperature dependence 
mentioned above, i.e., the highly-insulating state above $T_{\rm C}$ 
followed by a sudden drop below $T_{\rm C}$. 
There, the typical enhanced CMR effect is obtained 
by applying a small external magnetic field
\cite{Akahoshi2003}. 
Similar phenomena are also observed 
in $A_{0.55}$(Ca,Sr)$_{0.45}$MnO$_3$ 
\cite{Tomioka2002,Tomioka2003}
and in Cr doping in $A_{1/2}$Ca$_{1/2}$(Mn,Cr)O$_3$ 
\cite{Barnabe1997,Kimura1999,Moritomo1999,Katsufuji1999}. 
Another crucial observation is that 
strong fluctuations of charge and lattice orderings 
are observed in the highly-insulating state above $T_{\rm C}$, 
but they are largely suppressed below $T_{\rm C}$ 
\cite{Tomioka2003}.
This reentrant behavior appears to correlate with the temperature 
dependence of the resistivity. 
These experimental results strongly suggest that 
the disorder 
plays a significant role 
in the multicritical phenomena 
to induce the enhanced CMR effect. 

The purpose of this Contribution is 
to investigate the phase competition in the multicritical regime and 
the disorder effect theoretically. 
We will explore a comprehensive understanding of the enhanced CMR effect
on the basis of a cooperative-phenomenon picture 
rather than a single-particle one. 
In the next section \ref{sec:2}, we introduce 
an extended DE model to describe 
the phase competition, and 
briefly explain the numerical method. 
The results are presented in Sec.~\ref{sec:3}. 
We discuss the origin of CMR on the basis of our results in Sec.~\ref{sec:4}. 
We also compare our results with experimental data and 
other theoretical proposals. 
Sec.~\ref{sec:5} is devoted to summary and concluding remarks.

\section{Model and Method}
\label{sec:2}

In the present study, we consider a minimal model 
which captures the competition between FM and COI. 
We take account of the conventional DE interaction 
as a stabilization mechanism of FM state. 
As for COI, we consider one of the simplest mechanisms, 
i.e., the electron-phonon coupling to the breathing-type distortions
\cite{Verges2002}. 
The explicit form of the Hamiltonian reads
\cite{Motome2003} 
\begin{eqnarray}
H &=& -t \sum_{\langle ij \rangle \sigma} 
(c_{i\sigma}^{\dagger} c_{j\sigma} + {\rm H.c.}) 
- J_{\rm H} \sum_i \sigma_i^z S_i 
\nonumber
\\
&& -g \sum_i n_i Q_i + \frac12 \sum_i Q_i^2 
+ \frac{\lambda}{2} \sum_{\langle ij \rangle} Q_i Q_j 
+ \sum_i \epsilon_i n_i,
\label{eq:H}
\end{eqnarray}
where the summation with $\langle ij \rangle$ is taken over 
the nearest-neighbor sites $i$ and $j$, and 
the index $\sigma$ denotes the spin of conduction electrons. 
The first line of Eq.~(\ref{eq:H}) is for the DE part; 
the first term describes the kinetic energy of 
the single-band conduction electrons 
with the transfer integral $t$, and 
the second term denotes the ferromagnetic Hund's coupling 
between the conduction electron spin $\sigma_i$ and 
the localized spin $S_i$. 
For simplicity, we consider the limit of $J_{\rm H} \rightarrow \infty$ 
and the coupling of the Ising symmetry, i.e., $S_i = \pm 1$, 
which retains the essential physics of the DE ferromagnetism
\cite{Motome2001}. 
The first term in the second line of Eq.~(\ref{eq:H}) describes 
the electron-phonon coupling of the breathing type 
where $g$ is the coupling constant, 
$n_i = \sum_{\sigma} c_{i\sigma}^{\dagger} c_{i\sigma}$ 
is the local electron density, and 
$Q_i$ is the amplitude of the distortion at the site $i$. 
The next two terms denote the elastic energy of distortions. 
The latter term 
describes a cooperative aspect of the lattice distortion, where 
$\lambda$ is taken to be positive 
because a shrinkage (expansion) of a MnO$_6$ octahedron tends to 
expand (shrink) the neighboring MnO$_6$ octahedra. 
Lattice distortions are treated as classical objects for simplicity. 
The last term in Eq.~(\ref{eq:H}) is for the quenched disorder 
which couples to conduction electrons. 
In real materials, the alloying effect of $A$-site ions in $A$MnO$_3$ 
as well as of the Cr substitution into the Mn sites causes 
structural and electrostatic disorder, 
which modifies the on-site potential through the Madelung energy. 
Here, we mimic this by the random on-site energy $\epsilon_i$. 

We consider the model (\ref{eq:H}) on the two-dimensional (2D) square lattice 
in the half-doped case, i.e., $0.5$ electron per site on average. 
We set the half-bandwidth in the case of $J_{\rm H} = g = \epsilon_i = 0$ 
as an energy unit, i.e., $W = 4t = 1$. 
We take $\lambda = 0.1$ hereafter. 
We consider the binary-type distribution of the random potential, 
i.e., $\epsilon_i = \pm \Delta$. 

When $g$ is small (large bandwidth), the DE part is dominant and stabilizes FM 
at low temperatures. 
On the other hand, when $g$ is large (small bandwidth), 
the electron-phonon coupling becomes dominant and 
induces the checkerboard-type charge order (CO) with the wave number $(\pi, \pi)$. 
Hence, the competition between FM and COI is expected in our model (\ref{eq:H}) 
by controlling $g/W$. 
In the $A$-site ordered materials, 
the $A$-site substitution modifies mainly the bandwidth of conduction electrons 
through changes of the length and the angle of Mn-O-Mn bonds. 
The right (left) hand side of the phase diagram 
in Fig.~\ref{fig:phase diagram exp} 
corresponds to larger (smaller) bandwidth regime 
where FM (COI) is stabilized at low temperatures. 
In the $A$-site disordered materials, the situation is more complicated; 
the $A$-site substitution may change the bandwidth as well as 
the strength of the disorder. 
In the following, we investigate the model (\ref{eq:H}) 
by changing both $g$ and $\Delta$ systematically 
to understand the physics of the competing phases observed in experiments. 

We study thermodynamic properties of the model (\ref{eq:H}) 
by employing the Monte Carlo (MC) simulation. 
In the MC sampling, physical quantities are averaged 
for configurations of localized spins $\{ S_i \}$ 
and lattice distortions $\{ Q_i \}$ which are randomly generated 
by using the importance sampling technique. 
The MC weight is calculated 
by the diagonalization of the electronic part 
for a given configuration of $\{ S_i \}$ and $\{ Q_i \}$. 
In the presence of disorder, we take a quenched random average 
for different configurations of the on-site random potential 
$\{ \epsilon_i \}$. 
For the details of MC calculations, readers are referred to 
Ref.~\cite{Motome2003}. 
In the present system, since the competition between different orders as well as 
the spatial inhomogeneity due to the disorder is important, 
it is crucial to distinguish 
short-range correlations and long-range orders 
by applying the systematic finite-size scaling analyses
\cite{Motome2004}.

\section{Results}
\label{sec:3}

\subsection{Disorder Effect on Multicritical Phase Diagram}
\label{sec:3.1}

\begin{figure}
\centering
\includegraphics[width=11cm]{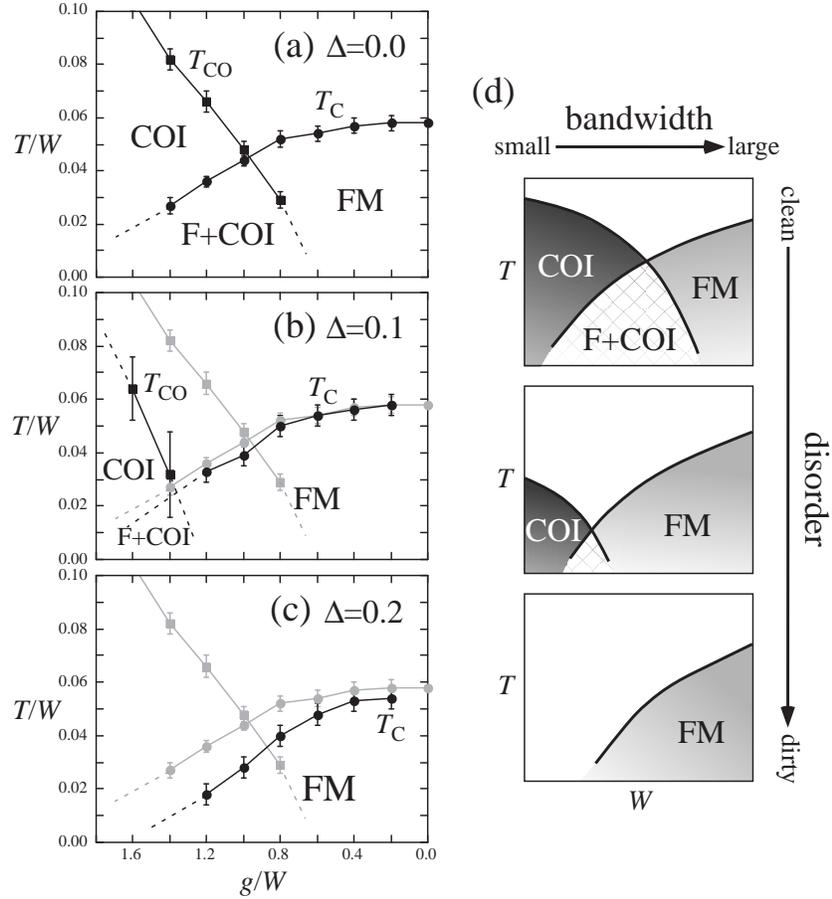}
\caption
{
(a)-(c) Phase diagrams of the model (\ref{eq:H}) 
for the change of the disorder strength $\Delta$. 
Circles (squares) denote the transition temperatures 
for the ferromagnetism (the charge ordering) 
estimated by using the finite-size scaling up to $8 \times 8$ sites. 
The lines are guides for the eyes. 
In (b) and (c), the results for the pure case (a)
are shown by gray symbols and lines for comparison. 
The change of the phase diagram is schematically summarized in (d). 
}
\label{fig:phase diagram}
\end{figure}

In the pure case without disorder ($\Delta = 0$), 
the model (\ref{eq:H}) shows the multicritical phase diagram 
as shown in Fig.~\ref{fig:phase diagram} (a). 
There, we have four different phases; 
the high-temperature para phase, 
FM phase in the small $g/W$ regime, 
COI phase in the large $g/W$ regime, and 
the coexisting phase of ferromagnetism and charge ordering (F+COI) in between. 
Note that the coexisting F+COI state is uniform and not phase-separated.
Thus, the phase diagram shows a tetracritical topology 
\cite{note:bicritical}.
From the systematic study of the density of states (not shown here), 
we find a metal-insulator transition at the phase boundary 
between FM and F+COI phases. 

When we introduce the disorder to the system, 
the multicritical phase diagram shows 
the systematic and drastic change 
as shown in Figs.~\ref{fig:phase diagram} (b) and (c). 
The FM phase is robust to the disorder 
showing a slight decrease of $T_{\rm C}$, 
whereas the COI phase is surprisingly unstable to the disorder. 
The critical temperature $T_{\rm CO}$ rapidly decreases 
with increasing the strength of disorder $\Delta$, and 
the COI phase finally collapses at $\Delta \sim 0.2W$ 
in the parameter range of the present MC calculations. 

The asymmetric change of 
the multicritical phase diagram for the disorder 
is schematically summarized in Fig.~\ref{fig:phase diagram} (d). 
In the following, we will explore the mechanism of this change 
to clarify the origin of the enhanced CMR.

\subsection{Fragility of Commensurate Insulator against Disorder}
\label{sec:3.2}

FM state is robust to the disorder 
because the DE ferromagnetism is stabilized 
by the kinetics of conduction electrons. 
$T_{\rm C}$ is proportional 
to the kinetic energy which gradually decreases with the disorder
\cite{Motome2003a}. 
On the contrary, we found that COI is surprisingly fragile against the disorder. 
The fragility is understood by the out-of-phase pinning phenomenon as follows. 

\begin{figure}
\centering
\includegraphics[width=11cm]{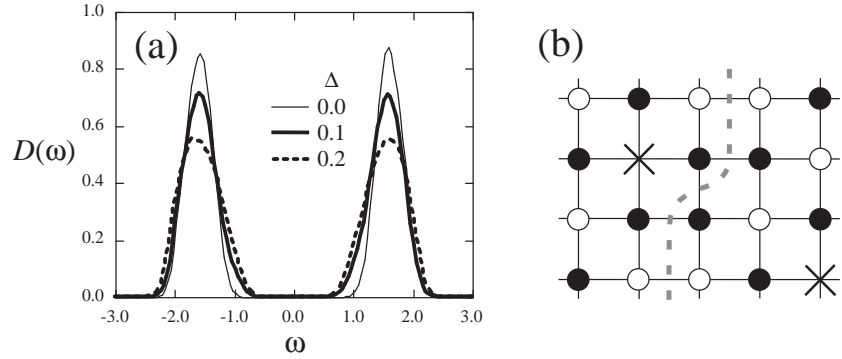}
\caption
{
(a) The density of states at $g = 1.6$ and $T = 0.016$ 
calculated for $8 \times 8$-sites systems. 
(b) A schematic picture of the out-of-phase pinning 
for the checkerboard-type charge ordering. 
White (black) circles denote electrons (holes), and 
crosses are the random pinning centers. 
The dashed gray curve shows an antiphase domain wall. 
}
\label{fig:DOS COI}
\end{figure}

Figure~\ref{fig:DOS COI} (a) shows the MC results of 
the density of states (DOS) in the COI regime. 
Even when the disorder destroys the long-range CO, 
the gap structure in DOS remains robust 
as shown in the figure. 
This means that short-range correlations survive and 
local lattice distortions persist to open the gap. 
Therefore, the fragility of the COI phase is not due to 
the rapid decrease of the amplitude of lattice distortions, 
but due to the disturbance of the phase of the commensurate ordering. 
The random potential to electrons is `a random field' 
to CO, and 
it easily pins the commensurate ordering pattern 
with introducing antiphase domain walls 
as schematically shown in Fig.~\ref{fig:DOS COI} (b).  

\subsection{Highly-Insulating State above $T_{\rm C}$}
\label{sec:3.3}

\begin{figure}
\centering
\includegraphics[width=6cm]{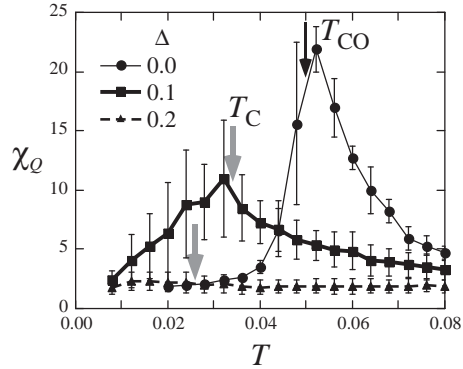}
\caption
{
Temperature dependence of the staggered susceptibility 
of the lattice distortion 
calculated at $g=1.0$ for $8 \times 8$-sites systems. 
The gray arrows indicate the values of $T_{\rm C}$ 
in the presence of disorder.  
}
\label{fig:kai}
\end{figure}

The out-of-phase pinning picture suggests that 
there remain short-range charge correlations and fluctuations 
in the region where the long-range CO is destroyed by the disorder. 
This is indeed the case. 
Figure~\ref{fig:kai} shows the temperature dependences of 
the susceptibility of the staggered lattice distortion 
which is calculated by fluctuations of the COI order parameter. 
Although the long-range CO is destroyed by the disorder and 
the diverging behavior at $T_{\rm CO}$ in the pure case is smeared out, 
fluctuations of charge and lattice orderings remain finite 
and are enhanced toward $T_{\rm C}$ 
even at a finite $\Delta$. 
This could be regarded as a reminiscence of 
the multicritical phenomenon in the pure case 
\cite{Murakami2003}. 

We note that the fluctuations are suppressed for large values of $\Delta$. 
We consider that for very strong disorder, 
the system may favor rather a static state with small COI clusters 
rather than the fluctuating state. 
This crossover will be discussed in Sec.~\ref{sec:4.3}. 

\begin{figure}
\centering
\includegraphics[width=11cm]{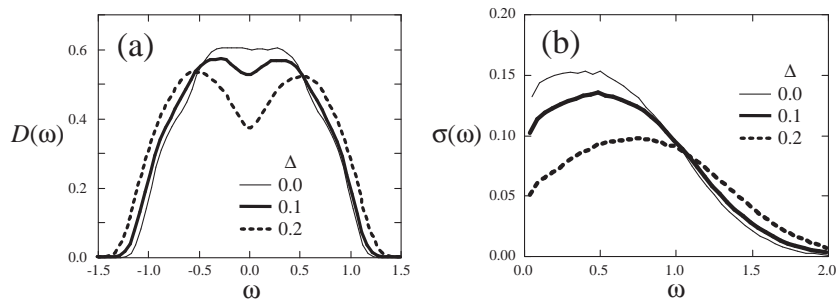}
\caption
{
(a) The density of states ($8 \times 8$ sites) and  
(b) the optical conductivity ($6 \times 6$ sites) 
at $g = 0.8$ and $T = 0.06$. 
}
\label{fig:highT}
\end{figure}

The remaining fluctuations of CO affect the electronic states 
in the high-temperature regime. 
Figures~\ref{fig:highT} show the MC results of 
(a) DOS and (b) the optical conductivity in this regime. 
As $\Delta$ becomes larger, 
the dip in DOS at the Fermi energy becomes deeper, and 
the low-energy weight of $\sigma(\omega)$ becomes smaller 
to develop a quasi-gap feature. 
Thus, the disorder induces fluctuations of CO, and 
tends to make the system insulating in the high-temperature regime. 

\subsection{Disorder-Induced Insulator-to-Metal Transition}
\label{sec:3.4}

\begin{figure}
\centering
\includegraphics[width=11cm]{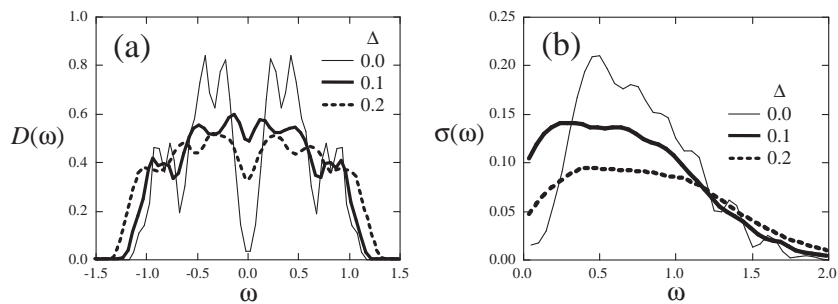}
\caption
{
(a) The density of states at $g = 0.8$ and $T = 0.016$ ($8 \times 8$ sites). 
(b) The optical conductivity at $g = 0.8$ and $T = 0.02$ ($6 \times 6$ sites). 
}
\label{fig:ItoM}
\end{figure}

In contrast to the enhanced insulating nature at high temperatures, 
we find a remarkable phenomenon at low temperatures 
in the multicritical regime, that is, 
the disorder-induced insulator-to-metal transition. 
Figures~\ref{fig:ItoM} show (a) DOS and 
(b) the optical conductivity 
close to the metal-insulator phase boundary. 
In the pure case ($\Delta=0$), DOS shows a finite gap and 
$\sigma(\omega)$ has a gap structure due to the long-range CO. 
When we introduce the disorder, the gap is suddenly filled and 
there appears a finite DOS at the Fermi energy. 
At the same time, the low-energy spectral weight of $\sigma(\omega)$ 
rapidly increases. 
These strongly indicate that the disorder induces 
the transition from the commensurate insulator to a metallic state. 
Such a phenomenon caused by the multicritical phase competitions 
is surprising and counter-intuitive 
because in general the disorder tends to localize electrons and 
to make the system more insulating. 

We note that the optical conductivity in the multicritical regime 
in Fig.~\ref{fig:ItoM} (b) as well as in Fig.~\ref{fig:highT} (b) 
shows drastic reconstruction in the wide energy range up to 
$\omega \sim W$ by a small disorder of only the order of $0.1W$. 
This is a consequence of the cooperative phenomena 
in the phase-competing regime.

\subsection{Brief Summary of Results}
\label{sec:3.5}

Here, we give a short summary of our results. 
MC simulation revealed that our model (\ref{eq:H}) exhibits 
the multicritical phenomenon due to the competition 
between FM and COI states, and 
the competing phases show the asymmetric response to the disorder. 
The COI phase is very unstable to the disorder 
while the FM phase is rather robust, which leads to 
the disorder-induced insulator-to-metal transition at low temperatures. 
In spite that the long-range CO is destroyed at low temperatures, 
the short-range correlations of CO are enhanced by the disorder 
at high temperatures, 
which tends to make the system more insulating. 
Thus, we obtain reentrant behavior, i.e., 
charge and lattice correlations are enhanced as decreasing temperature, but 
they are suppressed below $T_{\rm C}$. 
This is considered to be entropy-driven reentrant behavior; 
the short-range CO state has rather high entropy 
related to configurations of the antiphase domain walls, and
is favored at high temperatures due to the high entropy. 
We consider that 
this reentrant behavior is 
an origin of CMR phenomena as discussed in Sec.~\ref{sec:4.1}.

Our results elucidate an important role of 
the contrasting nature of the competing phases. 
The FM phase is stabilized by the kinetics of conduction electrons, and 
has a long-range ordering with the wave number ($0,0$). 
On the other hand, the COI phase is stabilized by 
a commensurate lattice ordering with the wave number ($\pi,\pi$). 
The former is robust but the latter is fragile 
against the out-of-phase pinning by disorder. 
The competition between the uniform metal and 
the commensurate insulator plays a central role.

\section{Discussions}
\label{sec:4}

\subsection{Origin of CMR}
\label{sec:4.1}

\begin{figure}
\centering
\includegraphics[width=11cm]{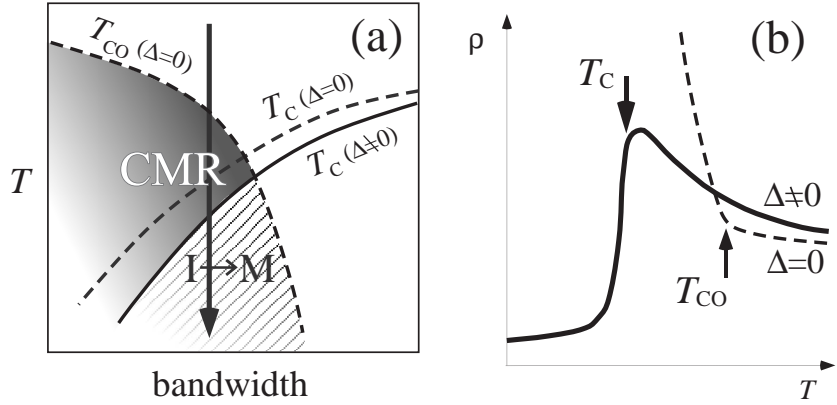}
\caption
{
(a) Schematic phase diagram of the model (\ref{eq:H}). 
Dashed curves denote $T_{\rm C}$ and $T_{\rm CO}$ 
in the absence of disorder, and 
the solid curve denotes $T_{\rm C}$ in the presence of disorder. 
See Fig.~\ref{fig:phase diagram} (d). 
The hatched area shows the region where 
the disorder-induced insulator-to-metal transition takes place. 
The shaded area denotes the relevant regime to the huge CMR effect 
where COI fluctuations are substantial and 
the highly-insulating state is realized. 
(b) Temperature dependences of the resistivity 
along the downward arrow in (a) suggested by the present study. 
Dashed and solid curves correspond to the cases 
without and with disorder, respectively. 
}
\label{fig:CMR}
\end{figure}

In this section, we discuss the origin of CMR suggested by our results. 
We found in Sec.~\ref{sec:3} 
the contrasting effects of disorder on the electronic states 
in the phase competing regime, i.e.,  
the insulator-to-metal transition at low temperatures and 
the enhanced insulating nature at high temperatures. 
The former occurs in the hatched area 
in the schematic phase diagram shown in Fig.~\ref{fig:CMR} (a). 
The latter is conspicuous in the shaded area in Fig.~\ref{fig:CMR} (a) 
where we found the enhanced fluctuations of charge and lattice orderings. 

The contrasting effects of disorder and 
the resultant reentrant behavior of charge and lattice correlations 
provide a key to understand 
the enhanced CMR effect observed in experiments. 
Schematic temperature dependences of the resistivity 
along the downward arrow in Fig.~\ref{fig:CMR} (a) 
are shown in Fig.~\ref{fig:CMR} (b), suggested 
by our numerical results of $\sigma(\omega)$ in Sec.~\ref{sec:3}. 
In the pure case with $\Delta = 0$, the resistivity sharply increases 
below the COI transition temperature $T_{\rm CO}$ 
as shown by the dashed curve in Fig.~\ref{fig:CMR} (b). 
When the disorder is introduced into the system, 
the high-temperature resistivity increases  
because the disorder enhances the insulating nature there. 
On the contrary, at a low temperature, 
we have the insulator-to-metal transition induced by the disorder, 
where the resistivity should show a sudden drop. 
Hence, we obtain the characteristic temperature dependence 
for $\Delta \neq 0$ as shown by the solid curve in Fig.~\ref{fig:CMR} (b). 
Note that the temperature dependence is in accord with 
that of charge and lattice fluctuations found in Fig.~\ref{fig:kai}, 
which illuminates the close relation 
between the resistivity and these fluctuations. 

The enhanced insulating state near above $T_{\rm C}$ is known to 
show a huge response to 
a small external magnetic field, i.e., the enhanced CMR 
as mentioned in Sec.~\ref{sec:1}. 
Such a huge response is expected in the shaded region in Fig.~\ref{fig:CMR} (a) 
where fluctuations of charge and lattice orderings 
are enhanced by the disorder. 
Thus, the enhanced CMR phenomena 
can be understood by the contrasting effects of the disorder 
in the phase competition between the uniform metal and 
the commensurate insulator. 

\subsection{Comparison with Experimental Results}
\label{sec:4.2}

Our results well reproduce the experimental results in CMR manganites 
of the $A$-site ordered/disordered materials
\cite{Nakajima2002,Akahoshi2003}, 
$A_{0.55}$(Ca,Sr)$_{0.45}$MnO$_3$ 
\cite{Tomioka2002,Tomioka2003}, 
and Cr-doped materials $A_{1/2}$Ca$_{1/2}$(Mn,Cr)O$_3$ 
\cite{Barnabe1997,Kimura1999,Moritomo1999,Katsufuji1999} 
in the following aspects;
\begin{itemize}

\item multicritical phase diagram in clean cases, 

\item asymmetric change of the multicritical phase diagram 
by introducing disorder, in particular, the fragility of COI phase, 

\item disorder-induced insulator-to-metal transition 
in the multicritical regime, 

\item highly-insulating state above $T_{\rm C}$ in the presence of disorder, 

\item remaining short-range correlations or fluctuations 
of charge and lattice orderings 
in the insulating state above $T_{\rm C}$
observed in Raman scattering and diffuse X-ray scattering experiments,

\item reentrant temperature dependence commonly observed in 
the resistivity and the charge and lattice fluctuations.

\end{itemize}
The agreement strongly indicates that 
our model (\ref{eq:H}) captures the essential physics of 
phase competing phenomena and 
disorder effects on them in CMR materials. 
We also note that our results give an understanding of 
the close relation between the charge fluctuations and CMR 
found in La$_{1-x}$Ca$_x$MnO$_3$ at $x \sim 1/3$
\cite{Dai2000,Adams2000}.

\subsection{Comparison with Other Theoretical Scenarios}
\label{sec:4.3}

Finally, we compare our scenario with 
other theoretical proposals for the mechanism of CMR.
As discussed in Sec.~\ref{sec:1}, 
CMR is a kind of localization-delocalization phenomenon 
at the ferromagnetic transition temperature $T_{\rm C}$. 
The low-temperature metallic phase is basically understood by 
the Zener's DE mechanism, and hence, 
the problem is how to understand the high-temperature insulating state 
above $T_{\rm C}$. 

Several proposals are based on the single-particle picture. 
One is the polaron scenario including both spin polaron 
\cite{Varma1996}
and JT polaron theories 
\cite{Millis1995,Roder1996}.
In this scenario, the high-temperature insulating state is 
described by the self trapping of electrons to form small polarons. 
Another scenario is the Anderson localization 
due to the quenched disorder 
\cite{Letfulov2001,Narimanov2002}. 
For both scenarios, it is difficult to answer the following questions 
on the general aspects of the enhanced CMR comprehensively:
\begin{itemize}

\item Why does the localized state at high temperatures become 
unstable at lower temperatures?

\item Why does the localization-delocalization transition always 
coincide with the ferromagnetic transition 
without fine tuning of parameters?

\item Why does the disorder cause the insulator-to-metal transition 
as observed in the $A$-site ordered manganites?

\item Why is CMR observed in wide region of the phase diagram, or 
why does the localization occur even at highly carrier-doped region?

\end{itemize}
Hence, the single-particle pictures 
appear to be insufficient to explain the enhanced CMR. 
In particular, they cannot explain the experimental fact that 
the `clean' system shows the multicritical phase diagram 
while the disorder induces the CMR phenomenon in the multicritical regime. 

On the contrary, 
our results based on the cooperative-phenomenon picture 
give a comprehensive understanding of 
CMR as a localization-delocalization phenomenon. 
Essential ingredients are the phase competition and the disorder, 
which are found in general in manganites and do not need 
fine tuning of parameters. 
Our theory answers why the delocalization occurs at $T_{\rm C}$ 
as well as why the disorder induces the insulator-to-metal transition. 

We note that there have been proposed 
other cooperative-phenomenon scenarios. 
One is the multicritical-fluctuation scenario
\cite{Murakami2003} 
in which enhanced fluctuations near the multicritical point play 
an important role in the insulating state above $T_{\rm C}$. 
The scaling law for the magnetization curve (the so-called Arrot plot) 
shows an evidence for this scenario. 
Although the scaling holds for a class of materials with weak    
disorder in which the multicritical point is at rather high  
temperature, it does not show good agreement with the others where 
the multicritical point is suppressed down to low temperatures by 
the disorder and the typical CMR is observed. 

Another proposal is the percolation scenario 
which explicitly takes account of the disorder. 
In this scenario, it is supposed that 
the disorder induces the percolated mixtures of FM and COI islands 
below $T^*$ which is the transition temperature $T_{\rm C}$ or 
$T_{\rm CO}$ in the absence of the disorder
\cite{Moreo1999,Burgy2001}. 
The resistivity is determined by the percolating path, and 
is sensitive to the pattern of the coexisting FM and COI regions 
which is easily changed by an external magnetic field. 
However, the theory is limited to a phenomenological level and 
does not give a microscopic explanation for 
the disorder-induced insulator-to-metal transition 
as well as the reentrant behavior of CO fluctuations.
Moreover, this rather static picture 
appears to be incompatible with recent experimental results 
which indicate a dynamically-fluctuating state 
just above $T_{\rm C}$
\cite{MoriPC}. 

Compared with above two scenarios based on the cooperative-phenomenon picture, 
our theory is considered to interpolate these two limiting pictures, i.e., 
it elucidates what happens in weak or moderate disorder regime 
in the phase-competing system. 
There, short-range correlations or 
dynamical fluctuations of charge and orbital orderings remain substantial 
and are relevant to the localization at high temperatures. 
In our numerical results in Sec.~\ref{sec:3}, 
we do not find 
any clear indication of the percolated cluster formation below $T^*$ 
\cite{Motome2003}, and 
such static phase separation may be 
relevant in rather strong disorder regime 
where CO fluctuations are largely suppressed 
as implied by Fig.~\ref{fig:kai}. 
Out theory, which describes the crossover 
from the clean to dirty limits, strongly suggests that 
the enhanced CMR occurs in the weak or moderate disorder regime 
with large fluctuations. 
Note that the unbiased numerical calculations play key roles 
to reveal the highly-nontrivial properties 
in the weak or moderate disorder regime .

\section{Summary and Concluding Remarks}
\label{sec:5}

In this Contribution, 
we discussed the phase competition 
between ferromagnetic metal and charge-ordered insulator 
and the role of the disorder 
by applying the Monte Carlo calculations 
to an extended double-exchange model. 
Highly nontrivial phenomena are revealed 
such as the disorder-induced insulator-to-metal transition 
as well as the entropy-driven reentrant behavior of charge-ordering fluctuations. 
Our results show good agreement 
with recent experimental results in CMR manganites, and 
give a comprehensive understanding of 
the origin of the enhanced CMR effect. 

There still remain many open problems. 
One is the role of other interactions 
which are neglected in our simplified model. 
One of the neglected elements 
is the orbital degree of freedom in the twofold $e_g$ orbitals 
which strongly couples with the JT lattice distortion. 
This is necessary, at least, to describe a complicated orbital ordering
in the COI phase
\cite{Yunoki1998}, 
and possibly introduces orbital fluctuations 
and enhances the insulating nature in the CMR regime 
as the fluctuations of charge and lattice orderings. 
Another omitted element is the AF superexchange interaction 
between localized spins. 
This is important to reproduce 
the complicated magnetic structure in the so-called CE phase 
as well as the A-type AF metallic phase, and 
possibly plays a substantial role in the glassy state 
at low temperatures. 
Recent theoretical study including these elements reported 
similar results to ours
\cite{Aliaga2003} 
although the numerical analysis is laborious and very limited 
for such a complicated model thus far. 

Another issue is the calculation in realistic 3D systems. 
The present calculations have been performed in 2D. 
The pinning effect due to the disorder is 
sensitive to the dimensionality, which should be carefully examined further. 
Moreover, in real materials, there may be some spatial correlation 
between the disorders, in other words, 
a long-range nature of the influence of disorder 
through the lattice strain effect or 
the cooperative effects of lattice distortions 
\cite{Burgy2004}. 
There, the orbital degree of freedom should play an important role 
through the Jahn-Teller coupling.
It is highly desired to examine such effects for realistic electronic models 
in the higher dimension. 

The direct calculation of the resistivity also remains 
as a hard task for numerical studies. 
There are several reports on the resistivity in small-size clusters 
which strongly indicate a huge response to the magnetic field 
\cite{Verges2002,SenPREPRINT}. 
To explain the enhanced CMR effect more quantitatively, 
further investigations are necessary 
including the development of theoretical tools  
to calculate the resistivity directly in larger-scale systems.

\section*{Acknowledgment}

The authors acknowledge Y. Tokura, Y. Tomioka, and E. Dagotto 
for fruitful discussions. 
This work is supported by Grants-in-Aid for Scientific Research and 
NAREGI Nanoscience Project from the Ministry of Education,  
Culture, Sports, Science, and Technology.

%
%

%
%



\printindex
\end{document}